
%
%
\input phyzzx
\font\seventeencp=cmcsc10 scaled \magstep3
\pubnum={IP--HET--95--6}
\date{September 1995}
\titlepage

\title{\seventeencp Path Integral Measure
via the Schwinger--Dyson Equations}

\author{A. R. Bogojevi\'c and D. Popovi\'c\footnote{}
{Email: bogojevic@castor.phy.bg.ac.yu\nextline
\phantom{Email: }popovic@castor.phy.bg.ac.yu}}

\address{Instite of Physics -- Belgrade\break
         Pregrevica 118, Zemun 11080, Yugoslavia}

\abstract
We present a way for calculating the Lagrangian path integral measure directly
from the Hamiltonian Schwinger--Dyson equations. The method agrees with the
usual way of deriving the measure, however it may be applied to all
theories, even when the corresponding momentum integration is not Gaussian.
Of particular interest is the connection that is made between the path
integral measure and the measure in the corresponding $0$-dimensional model.
This allows us to uniquely define the path integral even for the case of
Euclidean theories whose action is not bounded from bellow.

\endpage

\REFS\cs{J. C. Collins and D. E. Soper,
{\caps Large Order Expansion in Perturbation Theory}\nextline
{\it Ann. Phys.} {\bf 112} (1978)}
\REFSCON\david{F. David,
{\caps Phases of Large N Matrix Models and Non-Perturbative Effects
in 2d Gravity}\nextline
{\it Nucl. Phys.} {\bf b348} (1991)}
\REFSCON\ds{M. Douglas and S. Shenker,
{\caps Strings in Less Than 1d}\nextline
{\it Nucl. Phys.} {\bf b335} (1990)}
\REFSCON\bogojevic{A. R. Bogojevi\'c,
{\caps Unstable Field Theory}\nextline
Lectures at Danube Workshop '91}
\REFSCON\jeffreys{H. Jeffreys,
{\caps Asymptotic Approximations}\nextline
Oxford University Press, London (1962)}
\REFSCON\erdelyi{A. Erd\'elyi,
{\caps Asymptotic Expansion}\nextline
Dover Publications, New York (1956)}
\REFSCON\dingle{R. B. Dingle,
{\caps Asymptotic Expansions: Their Derivation and Interpretation}\nextline
Academic Press, New York (1973)}
\REFSCON\zwillinger{D. Zwillinger,
{\caps Handbook of Differential Equations}\nextline
Academic Press, New York (1989)}
\REFSCON\os{K. B. Oldham and J. Spanier,
{\caps The Fractional Calculus}\nextline
Academic Press, New York (1979)}

\chapter{Introduction}

The Schwinger--Dyson equations lie at the hart of the functional
formalism of quantum field theory. Given the complete set of basic amplitudes
({\it i.e.} Feymnan rules) for the propagator $\Delta_{ij}$, and the vertices
$\gamma_{ijk}$, $\gamma_{ijkl}, \ldots$ as well as the sources
$\jmath_i$, we can calculate the Green's functions of the theory:
$G_i$, $G_{ij}$, $G_{ijk},\ldots$. The conection between the basic
amplitudes and the Green's functions is given by an infinite set of
coupled equations called the Schwinger--Dyson equations. They are
simply the consequence of the basic linearity for the addition of amplitudes.
In their simplest form the SD equations are given in terms of the
generating functionals
$$
Z[\jmath]=\sum_{m=0}^\infty{i^m\over m!}\,G_{i_1i_2\ldots i_m}
\jmath_{i_1}\jmath_{i_2}\cdots\jmath_{i_m}\ ,\eqno(1.1)
$$
which generates the Green's functions, and
$$
\hat I[\phi]={1\over 2}\,\phi\Delta^{-1}_{ij}\phi_j+
{1\over 3!}\,\gamma_{ijk}\phi_i\phi_j\phi_k+
{1\over 4!}\,\gamma_{ijkl}\phi_i\phi_j\phi_k\phi_l+\ldots\eqno(1.2)
$$
which generates the Feynman rules. The Schwinger--Dyson equaions now
take the simple form reminiscent of the classical equations of motion
$$
\left({\partial \hat I\over\partial\phi_i}
\Bigm|_{\phi={1\over i}\,{\partial\over\partial\jmath}}+\jmath_i\right)
Z[\jmath]=0\ .\eqno(1.3)
$$
This is a linear (functional) differential equation for $Z[\jmath]$.
Note that the Fourier transform of (3) is just the Feynman path
integral representation of $Z[\jmath]$. The semi--classical limit of
$Z[\jmath]$ is dominated by configurations near
${\partial \hat I\over\partial\phi_i}=0$. From this we see that we
may write
$$
\hat I[\phi]=I[\phi]+{\hbar\over i}M[\phi]\ ,\eqno(1.4)
$$
where $I[\phi]$ is the classical action of the theory, and $M[\phi]$
is the measure term. Though this connection is beautiful this is as
far as the usual functional formalism takes us -- namely there is no
way to determine the measure term. The only way to do this is to make
connection with the operator formalism. From it we find an expression
for the generating functional in terms of a Hamiltonian path integral.
$$
Z[\jmath]=\int [dpdq]\exp\left({i\over\hbar}\int dt
\big(p\dot q-H(q,p)+\jmath q\big)\right)\ ,\eqno(1.5)
$$
Here the measure is trivial. The Lagrangian expressions, including
the corresponding measure, are obtained by doing the momentum path integral.

\medskip
It is the aim of this paper to provide an alternate way for
calculating the measure inside the functional formalism. To do this
we shall use the Hamiltonian form of the Schwinger--Dyson equations.
In this way we shall determine a differential equation that is
satisfied by the measure, and solve it for various instructive models.
In the process we shall learn about which boundary conditions one
must impose on (3) to uniquely pick out a solution. This will enable
us to also tackle unstable field theories: Euclidean theories whose
action is not bounded from bellow, or conversley Minkowski theories
whose energy is not bounded from bellow. A proto--typical theory is
Einstein gravity in Euclidean space. Several authors have looked
at unstable field theories\refmark{\cs,\david,\ds,\bogojevic}, and found
that the answer is an analytic extension of the path integral, where one
deforms the contour of integration of the path integral. The choice
of contour was dictated by the specific model. For example, in
[\david] David determined the contour for his matrix model approach
to strings from the requirement that his non--perturbative results
match the well--known perturbative string results. The nice thing
about the SD approach to the measure is that it uniquely picks out
which contour one should use.

\chapter{The Basic Formalism}

The generating functional written as a Hamiltonian path integral is
given by
$$
Z[\jmath ,k]=\int [dpdq]\exp\left({i\over\hbar}\int dt
\big(p\dot q-H(q,p)+\jmath q+kp\big)\right)\ ,\eqno(2.1)
$$
where we have for later convenience added a source term for the
momenta. The Schwinger--Dyson equations are easily derived from the identities
$$\eqalign{
0&=\int [dpdq]{\delta\over\delta q}\exp\left({i\over\hbar}\int dt
\big(p\dot q-H(q,p)+\jmath q+kp\big)\right)\cr
0&=\int [dpdq]{\delta\over\delta p}\exp\left({i\over\hbar}\int dt
\big(p\dot q-H(q,p)+\jmath q+kp\big)\right)\ .\cr}
$$
This gives us
$$
\eqalign{
\left(\dot P+{\partial H(Q,P)\over\partial Q}-\jmath \right)Z[\jmath ,k]&=0\cr
\left(\dot Q-{\partial H(Q,P)\over\partial P}+k\right)Z[\jmath ,k]&=0\ ,\cr}
\eqno(2.2)
$$
where we have introduced $P={\hbar\over i}{\delta\over\delta k}$, and
$Q={\hbar\over i}{\delta\over\delta \jmath }$. The above Schwinger--Dyson
equations look just like the classical Hamiltonian equations of
motion. The only difference is that we have the following non-zero commutators
$$
\eqalign{
[P,k]&={\hbar\over i}\cr
[Q,\jmath]&={\hbar\over i}\ .\cr}\eqno(2.3)
$$
Note that in this formalism $P$ and $Q$ commute.

We will now use the above equations to derive the Lagrangian path
integral measure. As an example let us look at a model whose Hamiltonian
is simply
$$
H(q,p)={1\over 2}p^2+V(q)\ .\eqno(2.4)
$$
In this case the SD equations read
$$
\eqalign{
&\left(\dot P+V'(Q)-\jmath \right)Z[\jmath ,k]=0\cr
&\left(\dot Q-P+k\right)Z[\jmath ,k]=0\ ,\cr}\eqno(2.5)
$$
Differentiating the second of these equations with respect to time,
and then adding this to $(5a)$ we get an equation for $Q$ alone
$$
\left(\ddot Q-V'(Q)-(\jmath -\dot k)\right)Z[\jmath ,k]=0\ .\eqno(2.6)
$$
The action for this model is
$I[q]=\int dt\big({1\over 2}\dot q^2-V(q)\big)$. By introducing
$J=\jmath -\dot k$ we may write (6) as
$$
\left({\delta I\over\delta Q}+J\right)Z[\jmath ,k]=0\ ,\eqno(2.7)
$$
which is just the Lagrange formalism Schwinger--Dyson equation.
Fourier transforming this we get
$$
Z[\jmath ,k]=\int [dq]\exp\left({i\over\hbar}
\int dt\big(\,{1\over 2}\,\dot q^2-V(q) + (\jmath -\dot k)q\big)\right)\ .
$$
We can now turn off the source for momenta. The generating functional
$Z[\jmath ]=Z[\jmath ,k=0]$ equals
$$
Z[\jmath ]=\int [dq]\exp\left({i\over\hbar}
\int dt\big(\,{1\over 2}\,\dot q^2-V(q)+\jmath q\big)\right)\ .\eqno(2.8)
$$
We have just derived the well known result that the path integral measure
is trivial for models whose Hamiltonian is of the simple form given
in (4).

\medskip
Now let us look at a bit more complicated example. We consider a
model with Hamiltonian given by
$$
H(q,p)={1\over 2}\,G^{-1}(q)p^2+V(q)\ .\eqno(2.9)
$$
The Hamiltonian SD equations are now
$$
\eqalign{
&\left(\dot P-{1\over 2}G^{-2}(Q)G'(Q)P^2+V'(Q)-\jmath \right)Z[\jmath ,k]=0\cr
&\left(\dot Q-G^{-1}(Q)P+k\right)Z[\jmath ,k]=0\ .\cr}\eqno(2.10)
$$
We may write the second equation as $PZ=G(\dot Q+k)Z$ and use this to
get rid of the $P$ terms in the first equation. Therefore
$$
P^2Z=P(G\dot Q+Gk)Z=(G\dot Q+Gk)PZ+[P,k]GZ=
\big((G\dot Q+Gk)^2+{\hbar\over i}\,G\big)Z\ ,
$$
as well as
$$
\dot PZ=\big(G'\dot Q(\dot Q+k)+G(\ddot Q+\dot k)\big)Z\ .
$$
Substituting this into $(10a)$, and setting $k=0$ we get
$$
\big(G\ddot Q+{1\over 2} G'\dot Q^2 + V'-{1\over 2}\,{\hbar\over i}\,(\ln G)'
-\jmath \big)Z[\jmath ]=0\ .\eqno(2.11)
$$
This equation can be written as
$$
\left({\delta \hat I\over\delta Q}+\jmath \right)Z[\jmath ]=0\ ,\eqno(2.12)
$$
where $\hat I=I+{\hbar\over i}M$. The first term is the classical action
$I[q]=\int dt\,\big(\,{1\over 2}G(q)\dot q^2-V(q)\big)$, while the
second term gives a contribution to the measure and is given by
$M=\int dt\,\ln\sqrt{G}$. Fourier transforming (12) gives us
$$
Z[\jmath ]=\int\prod_t\left(dq(t)\sqrt{G(q)}\right)
\exp\left(\,{i\over \hbar}\big(I+\int dt\,\jmath q\big)\right)\ ,\eqno(2.13)
$$
which again agrees with the standard derivation of the Lagrangian
path integral in which one performs the Gaussian momentum integration in the
Hamiltonian path integral.

\medskip
The generalization of the previous example to more variables gives us
the $\sigma$--model
$$
L={1\over 2}\,G_{\alpha\beta}(q)\dot q^\alpha\dot q^\beta\ .\eqno(2.14)
$$
The Hamiltonian is given in terms of the inverse metric
$G^{\alpha\beta}$, and equals
$H={1\over 2}\,G^{\alpha\beta}p_\alpha p_\beta$. The SD equations become
$$
\eqalign{
&\left(\dot P_\alpha+{1\over 2}G^{\gamma\delta}_{\ ,\alpha}P_\gamma P_\delta
-\jmath _\alpha\right)Z[\jmath ,k]=0\cr
&\left(\dot Q^\alpha-G^{\alpha\beta}P_\beta+k^\alpha\right)Z[\jmath ,k]=0
\ .\cr}\eqno(2.15)
$$
Just as in the previous example it is a simple exercise to get rid of
the $P$ terms and derive the Lagrangian SD equation. It may be
compactly written as
$$
\left({\delta I\over\delta Q^\alpha}-i\hbar
{1\over\sqrt{G}}\partial_\alpha\sqrt{G}+\jmath _\alpha\right)Z[\jmath ]=0
\ ,\eqno(2.16)
$$
where $G=\hbox{det }G_{\alpha\beta}$. The corresponding path integral
has the familiar form
$$
Z[\jmath ]=\int\prod_t\left(dq(t)\sqrt{G(q)}\right)
\exp\left(\,{i\over \hbar}\big(I+\int dt\,\jmath _\alpha q^\alpha\big)\right)
\ .\eqno(2.17)
$$
{}From these examples it is obvious that the generalization from
$1$-dimensional
field theory, {\it i.e.} quantum mechanics, to $d$-dimensional field
theory is trivial. The $d$-dimensional expressions just contain more
dummy labels.

\chapter{Non-Trivial Examples}

In this section we will look at models whose Hamiltonians are not quadratic in
$p$. To begin with we look at the Hamiltonian
$$
H={1\over 3}\,p^3\ .\eqno(3.1)
$$
Obviously the energy will not be bounded from bellow, however, let us
not worry about this for the moment. Later we will se that the SD
equations will be able to make a sensible theory out of (1). The
Hamiltonian SD equations are
$$
\eqalign{
&(\dot P-\jmath )Z=0\cr
&(\dot Q-P^2+k)Z=0\ .\cr}\eqno(3.2)
$$
The second of these equations may be written as $P^2Z=(\dot Q+k)Z$.
Now we are faced with a problem. In order to get rid of the $P$
dependence of the first SD equation we need to know how $P$ acts on
the generating functional. Instead of this we are given how $P^2$
acts on $Z$. If $P$ and $k$ commuted then the answer would be simply
$PZ=\sqrt{\dot Q+k}\ Z$. In fact, as we shall see, this indeed holds
when we take $\hbar\to 0$. From its definition we have
$P={\hbar\over i}\,{\delta\over\delta k}$, so that what we have is actually
$$
{\delta^2\over\delta k^2}Z[\jmath ,k]=-{1\over\hbar^2}\,(k+\dot Q)
Z[\jmath ,k]\ .
$$
Let us note that this is actually just an ordinary differential
equation: $\jmath $ is just a label, and at the same time $\dot Q$ is
just a constant as far as $k$ differentiaition is concerned. Writting
$C$ instead of $\dot Q$ we have
$$
{d^2\over dk^2}Z=-{1\over\hbar^2}\,(k+C)Z\ .\eqno(3.3)
$$
We don't realy need to solve this -- all we need is to find $PZ$.
Because of this we impose
$$
{dZ\over dk}={i\over \hbar}F(k)Z\ .\eqno(3.4)
$$
Differentiating this and using (3) we find that $F$ satisfies the
Riccati equation
$$
-i\hbar {dF\over dk}+F^2=k+C\ .\eqno(3.5)
$$
There are two general ways for dealing with Riccati equations. The
first is to write $F\propto {W'\over W}$ and choose the constant of
proportionality in such a way that $W$ obeys a linear differential
equation of second order. This is however just our starting equation
(3), so this doesn't help us. The second way to solve Riccati
equations leads to the general solution when any particular solution
is known. Again this is of no use since we know no obvious particular solution
of (5). Equation (5), however, does have a natural small parameter in it,
and we can find perturbative solutions, {\it i.e.} solutions written in
terms of a power series in $\hbar$. We write
$F=F_0+\hbar F_1+\hbar^2 F_2+\ldots$ Equation (5) now gives
$$
\eqalign{
&\qquad F_0^2=k+C\cr
&-iF_0'+2F_0F_1=0\cr
&-iF_1'+F_1^2+2F_0F_2=0\cr
&\qquad\qquad\cdots\cr}
$$
Choosing the $+$ sign for $F_0$ we get
$$
\eqalign{
F_0&=\sqrt{k+C}\cr
F_1&={i\over 4}\,(k+C)^{-1}\cr
F_2&={5\over 32}\,(k+C)^{-5/2}\cr
&\qquad\cdots\cr}
$$
This gives us
$$
PZ=\left((k+\dot Q)^{1/2}+{i\hbar\over 4}\,(k+\dot Q)^{-1}+
{5\hbar^2\over 32}\,(k+\dot Q)^{-5/2}+\ldots\right)Z\ .
$$
Differentiating this, substituting into the first SD equation and
setting $k=0$ we get
$$
\left(\,{1\over 2}\dot Q^{-1/2}\ddot Q-{i\over 4}\hbar\dot
Q^{-2}\ddot Q-{25\over 64}\hbar^2\dot Q^{-7/2}\ddot Q+\ldots -\jmath \right)
Z[\jmath ]=0\ .\eqno(3.6)
$$
This is simply
$$
\left({\delta\hat I\over\delta Q}+\jmath \right)Z[\jmath ]=0\ ,
$$
where
$$
\hat I=I+{i\over 4}\hbar\int dt\,\ln \dot q - {5\over 48}\hbar^2\int dt
\,\dot q^{-3/2}+\ldots\eqno(3.7)
$$
To one loop the path integral may be written as
$$
Z[\jmath ]=\int \prod_t \left(dq(t)\dot q^{-1/4}\right)
\exp\left({i\over\hbar}\big(I+\int dt\,\jmath q\big)\right)\ .\eqno(3.8)
$$

Perturbative solutions like (7) are nice -- if there is nothing
better arround. However, for this model, we know that the standard
treatment of the theory does not work since $H$ is not bounded from bellow.
Let us therefore look at equation (3) again. If we introduce
$$
x=-\hbar^{-2/3}(k+C)\ ,\eqno(3.9)
$$
then the equation simplifies to
$$
{d^2Z\over dx^2}=xZ\ .\eqno(3.10)
$$
This is Airy's differential equation. Equation (10) represents the
{\it Escherichia coli} in the field of asyptotic expansions, ({\it i.e.}
semi-classical expansions)\refmark{\jeffreys,\erdelyi,\dingle}.
The general solution for $x\in \bf C$ can be written as the Airy integral
$$
f(x)={1\over 2\pi i}\int_{C}\,e^{tx-{1\over 3}t^3}dt\ .\eqno(3.11)
$$
For $f(x)$ to converge, the integrand must vanish at the end-points.
The contour can't be closed because the integral of an analytic function over
such a contour vanishes. We thus have three topologicaly distinct
contours availible corresponding to the end points at infinity with phases
$-{2\pi\over 3}$, ${2\pi\over 3}$, and $0$. If we label these points
as $x_1$, $x_2$, $x_3$ then contour $C_{ij}$ goes from $x_i$ to $x_j$.
In addition we also have $C_{12}+C_{23}+C_{31}=0$, so that this gives
us two independent solution to (10). This is just right since the
Airy equation is of second order. The standard choices are the two
real independent solutions
$$
\eqalign{
\hbox{Ai}(x)&=f_{12}(x)\cr
\hbox{Bi}(x)&=i\big(f_{23}(x)-f_{31}(x)\big)\ .\cr}\eqno(3.12)
$$
Note that (10) is the Schwinger--Dyson equation for the $0$-dimensional
``path integral'' (11), {\it i.e.} for a theory with action
$I={1\over 3}t^3-tx$. This is an Euclidean path integral, however, in
$0$-dimensions there is no difference. By writting $s=it$ we get
$f(x)={1\over 2\pi}\int_C ds\,e^{i({1\over 3}s^3+sx)}$, which is the
corresponding Minkowski expression.

The Airy functions can readily be asymptotically expanded by the
method of steepest descent. The saddle points given by $I'=0$ are at
$t=\pm\sqrt{x}$. Paths of steepest descent are given by
$\hbox{\sl Im}(I)=\hbox{const}$. If we write $t=u+iv$, and look at
$x$ real and positive, then the paths of steepest descent passing through
the saddle points are $v=0$ and $v^2=3u^2-3x$. We have
$I''(\pm\sqrt{x})=\pm 2\sqrt{x}$, so that the left saddle point
contributes when going through it along $v^2=3u^2-3x$, while the right saddle
point contributes when we pass through it along $v=0$. We thus get
the asymptotic formulas
$$
\eqalign{
\hbox{Ai}(x)&\sim {1\over 2\sqrt{\pi}}x^{-1/4}e^{-{2\over3}x^{3/2}}\cr
\hbox{Bi}(x)&\sim {1\over\sqrt{\pi}}x^{-1/4}e^{{2\over 3}x^{3/2}}\ .\cr}
\eqno(3.13)
$$

If we choose $Z=\hbox{Bi}(x)$ then using (13) we get
$Z\propto (k+C)^{-1/4}e^{{2\over 3}(-)^{3/2}{1\over\hbar}(k+C)^{3/2}}$.
We have $\arg{x}\in [-\pi,\pi)$, so that $(-)^{3/2}=i$. Therefore we find
$$
{i\over Z}\,{dZ\over dk}=-{1\over
4}(k+C)^{-1}+{i\over\hbar}(k+C)^{1/2}\ ,
$$
hence
$$
PZ=\left((k+\dot Q)^{1/2}+{1\over 4}i\hbar (k+\dot Q)^{-1}\right)Z
\ .\eqno(3.14)
$$
This is in agreement with our perturbative result. The choice of the
$\hbox{Ai}(x)$ solution gives us a similar result, but with a wrong
sign in front of the classical part of (14). On the other hand
if we choose the solution $Z=\hbox{Ai}(x)+{1\over 2}\hbox{Bi}(x)$ then
we find
$$
PZ=\left((k+\dot Q)^{1/2}
\tanh\big({2\over 3}\,{i\over\hbar}(k+\dot Q)^{3/2}\big)
+{1\over 4}i\hbar (k+\dot Q)^{-1}\right) Z\ ,
$$
which doesn't look at all like our perturbative solution. The above
solution differs from the perturbative one by pieces that are smaller
than any power of $\hbar$.

We have seen that the naive expansion in $\hbar$ automatically picks
out one solution of (3). The correct proceedure is thus to solve
(10). To this we need to add additional physical input that tells us
which initial conditions to choose (or in the language of the path
integral which contour to choose). This multitude of solutions to the
SD equations is {\it always} present. For a theory whose action is
for example
$I=\int dx\,\big(\,{1\over 2}(\partial\phi)^2+{1\over 2}m^2\phi^2+
{1\over n!}g\phi^n\big)$, the SD equation is a linear (functional)
differential equation of $(n-1)$st order. There are thus $n-1$
independent solutions. We naively solve the SD equation by a (functional)
Fourier transform. However, in this way we choose a specific contour
--- the real axis. For $n$ even this is indeed one of the possible
solutions. In fact it is the correct one as we know from the operator
formalism. For $n$ odd the real axis is not one of the allowed
contours and we seem to have a problem. As we have seen in the simple
example of Airy functions there in fact is no problem -- we just have
to be careful in choosing the correct contours. What is the problem
in such theories is that the standard operator formalism does not
work, so we seem to lack a criterion that will tell us which of
the allowed contours to choose.

\medskip
There is a rather natural way around
this obstacle. We propose that the correct contour is the unique one
that has the correct semi--classical limit. Said another way -- we
should choose the contour that has the correct physics up to one loop.
Let us see what this means on the example of Airy functions. The
naive contour would be the real axis. It is wrong since the path
integral doesn't converge. However, one can still formally calculate its
asymptotic expansion. What we find is that it only gets a contribution from
the right saddle point $t=\sqrt{x}$.
The left saddle point doesn't contribute because in going along the
real axis it represents a maximum of the action, not a minimum. Now let
us look at the true solutions. $\hbox{Ai}(x)$ only sees the left
saddle point. In the direction of its contour this saddle point is a
minimum of the action, so everything is ok, however, this doesn't
agree with the imposed semi--classical results. On the other hand
$\hbox{Bi}(x)$ only sees the right saddle point. The contributions
from the left saddle point cancel for the two contours $C_{23}$ and $-C_{32}$.
Therefore, $\hbox{Bi}(x)$ has precisely the correct semi--classical behaviour.
It is easy to see that it is the unique such solution of the Airy
differential equation.

\medskip
We have calculated the measure for our model using the SD equations.
The way the measure is usually calculated is by performing the
momentum integration in the Hamiltonian path integral. Thus, what we
have in fact solved is
$$
\int [dp]e^{{i\over\hbar}\int dt(p\dot q-{1\over 3}p^3)}\ .\eqno(3.15)
$$
As we have seen the solution was given in terms of the Airy
differential equation. This is not surprising. The Airy integral is
simply the $0$-dimensional version of (15). In fact the relation is
stronger since (15) is an integration over $p$ of an expression that
doesn't contain derivatives. Therefore
$$
\int [dp]e^{{i\over\hbar}\int dt(p\dot q-{1\over 3}p^3)}=
\prod_t f\big(-\dot q(t)\big)\ .\eqno(3.16)
$$
Now we come to an important point -- the choice of contour of the
$0$-dimensional integral completely determines the path integral (15). We
therefore need to use the $\hbox{Bi}(x)$ Airy function in (16). Once
we do this we get (to one loop)
$$
Z[\jmath ]=\int \prod_t\left(dq(t) \dot q^{-1/4}\right)e^{{i\over\hbar}I}\ ,
$$
which is precisely what we had before.

\medskip
The next example that we look at illustrates another novel aspect of
the SD approach to the measure. We will look at a model with Lagrangian
$$
L={1\over 3}\,\dot q^3\ .\eqno(3.17)
$$
The Hamiltonian one gets has momenta to a non-integer power
$$
H={2\over 3}\,p^{3/2}\ .\eqno(3.18)
$$
The SD equations are now
$$
\eqalign{
&(\dot P-\jmath )Z=0\cr
&(\dot Q-P^{1/2}+k)Z=0\ .\cr}\eqno(3.19)
$$
Equation (19b) is in fact an example of a so--called extra--ordinary
differential equation\refmark{\zwillinger,\os}, {\it i.e.} one that containes
derivatives to a fractional power. There are several ways to make sense of
such equations the simplest of which is by using Laplace transforms.
The Laplace transform is given by
$$
L\big(f(t)\big)=\int_0^{+\infty}dt\,f(t)e^{-ts}\ ,\eqno(3.20)
$$
and its inverse is
$$
L^{-1}\big(g(s)\big)={1\over 2\pi i}\,\int_{C-i\infty}^{C+i\infty}
ds\,g(s)e^{ts}\ .\eqno(3.21)
$$
{}From (20) we easily find the Laplace transform of an $n$-th derivative
to be
$$
L\big(f^{(n)}(t)\big)=sL\big(f^{(n-1)}(t)\big)-f^{(n-1)}(0)\ .\eqno(3.22)
$$
Iterating this $m$ times and then setting $n=m$ we get
$$
L\big(f^{(n)}(t)\big)=s^nL\big(f(t)\big)-f^{(n-1)}(0)
-sf^{(n-2)}(0)-\ldots -s^{n-1}f(0)\ .\eqno(3.23)
$$
It is convenient to define $f^{(n)}(t)$ for negative $n$'s. The
inverse of a derivative is an integral, so we may take
$f^{(-1)}(t)=\int_a^t du\,f(u)$. We can uniquely specify this by
imposing $f^{(-1)}(0)=0$, which gives $a=0$. Similarly we choose
$$
f^{(n)}(0)=0\quad\hbox{for}\quad n\le -1\eqno(3.24)
$$
Now we shall take (23) and (24) to be valid for all real values of
$n$. For example for $n=-1$ we get
$$
L\big(f^{(-1)}(t)\big)={1\over s}\,L\big(f(t)\big)\ .
$$
We can now use the inverse Laplace transform to see that what we get
precisely agrees with the definition of $f^{(-1)}(t)$ given above. For
$n={1\over 2}$ equation (23) gives
$$
L\big(f^{(1/2)}(t)\big)=\sqrt{s}\,L\big(f(t)\big)-f^{(-1/2)}(0)\ .\eqno(3.25)
$$
The inverse Laplace transform of this gives us our definition of
${d^{1/2}\over dt^{1/2}}$. For example, one can easily show that
${d^{1/2}\over dt^{1/2}}{1\over\sqrt{t}}=0$. From this example it is
obvious that in general
${d^{1/2}\over dt^{1/2}}{d^{1/2}\over dt^{1/2}}f\ne {d\over dt}f$.
In fact, as is shown in reference [\os] we have
${d^{1/2}\over dt^{1/2}}{d^{1/2}\over dt^{1/2}}f={d\over dt}f+Gx^{-3/2}$,
where the constant $G$ is determined via consistency conditions. For
example, for our previous example we have $G=1/2$.

\medskip
We are now finished with this mathematical asside, and
are ready to face equation (19b), which may be written as
${d^{1/2}\over dk^{1/2}}Z=\left(\sqrt{i\over\hbar}k+
\sqrt{i\over\hbar}C\right)Z$. If we set $k+C=i\hbar^{1/3}\,t$,
then this simplifies to
$$
\left({d\over dt}\right)^{1\over 2}\,Z=tZ\ .\eqno(3.26)
$$
Laplace transforming this and using (25) we get
$$
\sqrt{s}L(Z)-D=L(tZ)=-{d\over ds}L(Z)\ ,\eqno(3.27)
$$
where we have set $D=Z^{(-1/2)}(0)$. Introducing $L(Z)=F(s)$ allows us
to write the previous equation as
$$
{dF\over ds}=D-\sqrt{s}\,F(s)\ .\eqno(3.28)
$$
This is readily solved for $D=0$, where we find
$$
F(s)=E\,e^{-{2\over 3} s^{3/2}}\ ,\eqno(3.29)
$$
for constant $E$. The $D\ne 0$ equation has the same solution, only $E$
becomes a function $E(s)=D\int^s du\,\exp\left({2\over 3}u^{3/2}\right)+
\hbox{const }$. Although this is solved in quadratures we can't go any
further because we can't solve the above integral. Therefore, we will
continue working with the $D=0$ solution -- note that this choice picks
out a {\it specific} solution of (26). Taking the inverse Laplace
transform of (29) gives
$Z=E\,{1\over 2\pi i}\int_{C-i\infty}^{C+i\infty}ds\,e^{ts-{2\over 3}s^{3/2}}$.
This can be readily asymptotically expanded and one gets
$E\,{1\over 2\pi}e^{{1\over 3} t^3}\int_{-\infty}^{+\infty}dx\,e^{{1\over 4}
t^{-1}x^2}$.
This integral converges for $\hbox{\sl Re}(t)<0$, which is indeed the
case since $t$ is pure imaginary (with infinitesimal negative real part)
when $k$ is real. Therefore we have $Z\propto \sqrt{-t}e^{{1\over 3} t^3}$.
Going back to the original variables we after setting $k=0$
$$
Z\propto \sqrt{\dot q}\,e^{{i\over\hbar}{1\over 3}(\dot q)^3}\ .\eqno(3.30)
$$
Note that this is precisely what we get by doing the corresponding
momentum path integral -- once we choose the correct contour. By steepest
descent we find
$$
\int [dp]e^{{i\over\hbar}\int dt(p\dot q-{2\over 3}p^{3/2})}=
\prod_t\int dp(t)e^{{i\over\hbar}(p\dot q-{2\over 3}p^{3/2})}
\propto\prod_t\sqrt{\dot q}\,e^{{i\over\hbar}\int dt\,{1\over s}\dot
q^3}\ .
$$
This is precisely what we got in (30).

\medskip
Having derived these results the right way, we will now give a fast,
though formal derivation. We start from
$$
P^{1/2}Z=(\dot Q+k)Z\ .\eqno(3.31)
$$
We next multiply both sides with $P^{1/2}$. Remember
$P^{1/2}P^{1/2}Z=PZ+Gk^{-3/2}$, so we get
$$
PZ+Gk^{-3/2}=P^{1/2}(\dot Q+k)Z=(\dot Q+k)^2Z+ [P^{1/2},k]Z\ .
$$
Using the realtion $[V(P),k]=-i\hbar {dV\over dP}$, which is strictly
only valid for analytic functions $V(P)$, we get
$$
PZ+Gk^{-3/2}=(\dot Q+k)^2Z-i\hbar P^{-1/2}Z\ .\eqno(3.32)
$$
{}From (31) to order $\hbar^0$ we immediately get $P^{-1/2}Z=(\dot Q+k)^{-1}Z$,
hence to one loop we have
$$
PZ+Gk^{-3/2}=\left((\dot Q+k)^2-{1\over 2}i\hbar (\dot Q+k)^{-1}\right)Z
\ .\eqno(3.33)
$$
Differentiating this with respect to time, and setting $k=0$ we find
$$
\dot PZ=\left(2\ddot Q\dot Q+{1\over 2}i\hbar
\ddot Q\dot Q^{-2}\right)Z=0\ .\eqno(3.34)
$$
Note that we had to choose $G=0$ in order to get a finite result --
in fact that is our consistency condition. The above is precisely what we got
previously.

\chapter{Conclusion}

As we have seen, the SD equations offer us a new way to calculate the
measure for the Lagrangian path integral. All of our examples
concerned quantum mechanical systems, but the generalization to field
theory in more then one dimension is trivial. What is not trivial,
when one tackles full--fledged field theory, is how to deal with
gauge symetries, and anomalies. Therefore, it will be interesting to extend
this work to the treatment of gauge theories, and re--derive the
measures obtained by Faddeev--Popov and Batalin--Vilkovisky. Another
direction one should go is to try and cast the differential equation for
the measure not in terms of the Hamiltonian (as in this paper), but solely
in tems of the Lagrangian. Doing this will enable us to complete what Dirac
and Feynman have started: To define a complete quantum theory in terms of
the Lagrangian, {\it i.e.} the action.

\medskip
\ACK

We would like to thank Milutin Blagojevi\'c for many helpfull discussions.

\refout
\end